\documentclass[prd,twocolumn,superscriptaddress,altaffilletter,amssymb,showpacs,nofootinbib]{revtex4}
\usepackage[dvips]{graphicx}
\usepackage{amsmath}

\newcommand{\be}{\begin{equation}}
\newcommand{\ee}{\end{equation}}
\newcommand{\bea}{\begin{eqnarray}}
\newcommand{\eea}{\end{eqnarray}}

\newcommand{\vphi}{\varphi}

\begin{document}

\title{Matter Couplings which are Compatible with a Fundamental Theory of Physics}

\author{Israel Quiros}\email{iquiros6403@gmail.com}\affiliation{Departamento de Matem\'aticas, Centro Universitario de Ciencias Ex\'actas e Ingenier\'ias, Universidad de Guadalajara, Guadalajara, CP 44430, Jalisco, M\'exico.}

\date{\today}

\begin{abstract}
A criterion to seek for compatibility of any given matter coupling with a fundamental theory of physics -- assumed here to be the string theory -- is developed. We call this criterion as ``compatibility check'', and apply it to test different kinds of matter coupling which are familiar in cosmological scenarios. The so called ``chameleon effect'' -- useful to overcome the stringent constraints on additional non gravitational interactions -- is reviewed by relaxing several assumptions which are commonly found in the bibliography in order to explore the chameleon behavior of self-interacting scalar fields. Although the results of this study are far from being conclusive, these (i) suggest that matter couplings other than the exponential (dilaton) are very unlikely to arise in a cosmological setup, and (ii) pose reasonable doubts on the validity of the chameleon picture as we know it.
\end{abstract}

\maketitle

\section{Introduction}

Scalar fields have played a central role in cosmological evolution modeling designed to describe not only early inflation but, also, the present inflationary stage of the cosmic expansion \cite{de-models, wands}. The energy density associated with the scalar field's self-interaction potential is not diluted by the expansion of the Universe, so that the scalar field $\vphi$ behaves as an effective cosmological constant, responsible for the inflationary stage. 

In general it is expected that the scalar field may couple explicitly to ordinary matter unless some special symmetry prevents or suppresses the coupling \cite{amendola, carroll}. Non minimal coupling of matter is sustained by a fundamental theory of physics: string theory. An illustration can be the type IIB theory \cite{string-iib} which, in the Einstein's frame, is given by the following dimensionally reduced (dual) string effective action \cite{wands-rev}:

\bea &&S_*=\int_{\cal M} d^4x\sqrt{|g|}\left[R-\frac{1}{2}(\nabla\vphi)^2-\frac{1}{2}\,e^{2\vphi}(\nabla\sigma_1)^2\right.\nonumber\\
&&\;\;\;\;\;\;\;\;\;\;\;\;\;\;\;\;\;\;\;\;\;\;\;\;\;\;\;\;\;\left.-\frac{1}{2}\,e^\vphi\,e^{-\sqrt{3}\beta_0}(\nabla\sigma_2)^2+...\right],\label{string-iib-action}\eea where $(\nabla\chi)^2\equiv g^{\mu\nu}\nabla_\mu\chi\nabla_\nu\chi$, $\vphi$ stands for the dilaton, $\sigma_1$, $\sigma_2$ are the axion (pseudo-scalar) fields related with the three-form-fields, the dots account for terms potentially related with other scalar fields, and we have considered frozen radius of the internal space, which is parametrized by the modulus field $\beta=\beta_0$. As seen from (\ref{string-iib-action}), the terms related with the three-form matter fields are directly coupled to the dilaton through exponential couplings of different strength $e^\vphi$, and $e^{2\vphi}$, respectively.

In the cosmological context consideration of non minimal coupling or, alternatively, of additional non gravitational interactions between the dark matter and the dark energy has been repeatedly invoked to get late-time matter-scaling attractors which allow accelerated expansion \cite{amendola, interaction, brasileiros, luis-roy}. Besides, it is intuitive to imagine that, at least in principle, the different components of the cosmic mixture may exchange energy-momentum. In this regard, several more or less physically motivated kinds of matter coupling have been explored \cite{coley, turner, malik, cen, chen, saridakis}. Although there exist stringent constraints on additional non gravitational interactions coming from five-force experiments \cite{5-force}, and from equivalence principle testing \cite{will, baessler}, given the yet unknown nature of the cold dark matter and of the dark energy, it is possible to hypothesize a certain exchange of energy-momentum between these constituents of the cosmic substratum \cite{jakubi}.

A way to avoid the tight constraints on additional non gravitational interactions \cite{5-force, will, baessler} can be through the so called ``chameleon'' effect \cite{khoury, cham-brax-prd, cham-tsujikawa-jcap, cham-takami-prd, cham-salehi-prd, cham-polarski-prd, cham-banerjee-prd}. According to this effect, the scalar field's effective self-interaction potential is modified by the background (surrounding) energy density: At cosmological scales where the background energy density is vanishingly small, the effective mass of the scalar field is also vanishingly small and it modifies the cosmic dynamics, meanwhile, at local astrophysical scales the surrounding energy density is much bigger leading to a also large scalar field's effective mass so that the Yukawa-like term does not modify the Newtonian potential in any appreciable way. Regrettably, in order to get such a nice result a great deal of simplifying assumptions is usually undertaken. Among these assumptions we can list the following: (i) a single matter species considered \cite{khoury, cham-brax-prd, cham-tsujikawa-jcap, cham-takami-prd, cham-salehi-prd, cham-polarski-prd, cham-banerjee-prd}, non-relativistic most times (see, however \cite{cham-tsujikawa-jcap} for consideration of relativistic backgrounds) or, if a number of matter species is assumed -- the chameleon considered apart -- then, additional non gravitational interaction is assumed only with the chameleon field but not among the different matter species, (ii) consideration of self-interaction potentials of the runaway type only, which should be monotonically decreasing functions and should fulfill a number of assumptions on the asymptotic behavior of the ratio of their derivatives \cite{khoury, cham-polarski-prd}, (iii) geometry approximated by Minkowski space \cite{khoury, cham-takami-prd, cham-polarski-prd}, and (iv) a certain ``energy density which, in a cosmological setting would be conserved'' in the Einstein's frame, is commonly considered \cite{khoury}, instead of the energy density which is measured by co-moving observers, or the one which is conserved in the Jordan's frame. It is our opinion that, while the first assumption above might be reasonable, the remaining ones are unjustified and may lead to misleading results. 

In this paper we shall review the chameleon effect after relaxing several of the assumptions above but, before that, we shall explore a much serious problem: can be there any bound to the kind of matter coupling (if any) which could be found in nature? or, in other words; can be  admissible any imaginable type of matter coupling -- with a more or less sound physical basis -- as a (potential) model of additional non gravitational interactions of matter? Perhaps this question underlies a deeper problem of physics, and finding a final answer might take a huge amount of work and effort. Notwithstanding, in the present paper we shall make a modest step towards a better understanding of this issue by exploring, within a cosmological setting, to which extent several well known kinds of matter coupling are compatible with a fundamental theory of physics. Here we assume string theory to be such a fundamental theory.

We want to underline that in the present paper we shall be interested more in the cosmological/gravitational side of the matter coupling than in its particle side, i. e., we shall be focusing in coupling of macroscopic matter fields which can be modeled as perfect fluids. Couplings of individual quantum fields (fermions) to pseudo-scalar fields, etc., are behind the scope of this paper and will not be considered.

The paper has been organized in the following way. In the next section we remark on the only kind of matter coupling which is straightforwardly derived from the string effective theory: the exponential dilaton coupling. Other kinds of coupling which are useful within a cosmological setting are exposed in Sec. \ref{other-coupling}. In Sec. \ref{consistency} a criterion to seek for compatibility of any given matter coupling with a fundamental theory of physics -- assumed here to be the string theory -- is developed. We call this criterion as ``compatibility check'', and apply it to test different kinds of matter coupling which are familiar in cosmological scenarios. In section \ref{chameleon-coupling} we review the so called chameleon effect -- useful to overcome in an elegant way the stringent constraints on additional non gravitational interactions -- by relaxing several assumptions which are commonly used in the bibliography in order to expose the chameleon behavior of self-interacting scalar fields. The results of the paper are discussed in section \ref{discussion} while brief conclusions are given in section \ref{conclusion}.

\section{Remarks on Dilaton's Coupling to the Matter Degrees of Freedom}\label{dilaton-coupling}

The idea that the gravitating matter fields may interact in an additional non-gravitational way finds a natural action principle formulation within superstring effective theory, where the dilaton couples directly to matter \cite{wands-rev}. In the string frame the dimensionally reduced dilaton-graviton action can be written as $$S_{dg}=\frac{1}{2}\int_{\cal M}d^4x\sqrt{|g|}e^{-\phi}\left[R-\omega(\nabla\phi)^2-2e^{-\phi}V(\phi)\right],$$ where $\phi$ is the dilaton field, $\omega$ is the Brans-Dicke coupling parameter which quantifies the strength of the interaction between the Brans-Dicke field $e^{-\phi}$ and gravity,\footnote{For the string effective action, after dimensional reduction on a $d$-torus, $\omega=-1$ \cite{wands-rev}.} and we have considered an arbitrary self-interaction potential $e^{-\phi}V(\phi)$ for the dilaton. One may consider also a matter piece of action $$S_m=\int_{\cal M}d^4x\sqrt{|g|}{\cal L}_m(\mu,\nabla\mu,g_{\mu\nu}),$$ where $\mu$ stands for the matter fields. Under a conformal rescaling of the metric, jointly with a rescaling of the dilaton:

\be g_{\mu\nu}\rightarrow e^\phi g_{\mu\nu},\;\phi\rightarrow\beta\,\vphi,\;\beta=\frac{1}{\sqrt{\omega+3/2}},\label{conf-t}\ee the above actions transform into

\bea &&S_{dg}\rightarrow S_{dg}^E=\frac{1}{2}\int_{\cal M}d^4x\sqrt{|g|}\left[R-(\nabla\vphi)^2-2V(\vphi)\right],\nonumber\\
&&S_m\rightarrow S_m^E=\int_{\cal M}d^4x\sqrt{|g|}e^{2\beta\vphi}{\cal L}_m(\mu,\nabla\mu,e^{\beta\vphi}g_{\mu\nu}),\label{ef-actions}\eea where $S^E$ means that given quantities and operators within the action are expressed in terms of the conformal Einstein's frame (EF) metric. Notice that the matter fields do not couple to the EF metric but to the string frame one $e^{\beta\vphi}g_{\mu\nu}$.\footnote{As mentioned in the introduction, according to the spirit of string effective theory, the different matter fields may couple with different strength to the dilaton: $$e^{2\beta\vphi}{\cal L}_m(\mu,\nabla\mu,e^{\beta\vphi}g_{\mu\nu})\rightarrow\sum_i e^{2\beta_i\vphi}{\cal L}_m(\mu_i,\nabla\mu_i,e^{\beta_i\vphi}g_{\mu\nu}),$$ where the sub-index $i$ labels the different matter species.} 

Given the boundary term \cite{boundary} $S_{\cal B}=2\int_{\partial{\cal M}}d^3x\sqrt{|h|}\,K$, where $h_{\mu\nu}$ is the metric induced on the boundary $\partial{\cal M}$, and $K$ is the extrinsic curvature of the boundary hypersurface, then, extremising the action $S_{\text{tot}}=S_{dg}^E+S_m^E+S_{\cal B}$ with respect to any field variations that vanish on the boundary $\partial{\cal M}$, yields to the following field equations:  

\bea &&G_{\mu\nu}=T^{(m)}_{\mu\nu}+T^{(\vphi)}_{\mu\nu},\;\nabla^2\vphi=\frac{dV}{d\vphi}-\frac{\beta}{2}\,T^{(m)},\label{ef-feqs}\eea where $\nabla^2\equiv g^{\mu\nu}\nabla_\mu\nabla_\nu$, $$T^{(\phi)}_{\mu\nu}=\vphi_{,\mu}\vphi_{,\nu}-\frac{1}{2}g_{\mu\nu}(\nabla\vphi)^2-g_{\mu\nu}V(\vphi),$$ is the dilaton's stress-energy tensor, while $$T^{(m)}_{\mu\nu}=-\frac{2}{\sqrt{|g|}}\frac{\partial(\sqrt{|g|}e^{2\beta\vphi}{\cal L}_m)}{\partial g^{\mu\nu}},$$ is the stress-energy tensor for the matter degrees of freedom [$T^{(m)}=g^{\mu\nu}T^{(m)}_{\mu\nu}$]. Due to the Bianchi identities and to equations (\ref{ef-feqs}), the following continuity equation is obtained for the matter degrees of freedom:

\be\nabla^\mu T^{(m)}_{\mu\nu}=\frac{\beta}{2}\,\vphi_{,\nu}\,T^{(m)}.\label{m-cont-eq}\ee 

We see that the stress-energy is not separately conserved for each of the interacting species but, instead, it is the total stress-energy density which is conserved: $\nabla^\mu \left(T^{(m)}_{\mu\nu}+T^{(\vphi)}_{\mu\nu}\right)=0$, where we have taken into account that 

\be\nabla^\mu T^{(\vphi)}_{\mu\nu}=\left(\nabla^2\vphi-\frac{dV}{d\vphi}\right)\vphi_{,\nu}=-\frac{\beta}{2}\,\vphi_{,\nu}\,T^{(m)}.\label{vphi-cont-eq}\ee

The source term of the continuity equations (\ref{m-cont-eq}), (\ref{vphi-cont-eq}) is properly the interaction term which accounts for the additional non-gravitational interaction between the dilaton $\vphi$ and the matter degrees of freedom:

\bea &&\nabla^\nu T^{(m)}_{\nu\mu}=-Q_\mu,\;Q_\mu=-\frac{1}{2}\beta\,T^{(m)}\vphi_{,\mu}.\label{i-tensor}\eea 

In a cosmological setting,\footnote{Here we shall consider Friedmann-Robertson-Walker (FRW) spacetimes with flat spatial sections, depicted by the line-element $$ds^2=-dt^2+a^2(t)\delta_{ik}dx^idx^k,\;\;\;i,k=1,2,3.$$} since $\vphi_{,\mu}=(\dot\vphi,\vec{0})$, and assuming dust matter for which $T^{(m)}=-\rho_m$, one has

\bea Q_\mu=(Q,\vec{0}),\;Q=\beta\,\dot\vphi\,\rho_m/2.\label{string-i-term}\eea This is, strictly speaking, the only form of the non-gravitational coupling of matter which is derived from a fundamental theory: string theory \cite{amendola, wetterich}. 

For a more general form of the coupling in the matter action (\ref{ef-actions}); $e^{2\beta\vphi}\,{\cal L}_m\rightarrow\chi^2(\vphi)\,{\cal L}_m$, one obtains an interaction term of the following form:

\bea Q=\frac{1}{2}\frac{d\ln\chi}{d\vphi}\,\dot\vphi\,\rho_m.\label{lag-i-term}\eea 

Although the exponential coupling [$e^{2\beta\vphi}\,{\cal L}_m$] is the one which can be straightforwardly derived from a fundamental theory of physics, the coupling (\ref{lag-i-term}) represents the most general Lagrangian formulation of an additional non-gravitational interaction of matter with a scalar field.

\section{Other couplings of matter}\label{other-coupling}

In general non gravitational interactions of matter have proven useful to overcome the coincidence problem in the context of general relativity models of dark energy \cite{amendola, brasileiros, luis-roy, pavon, gumjudpai, cai, quiros, chen, saridakis, bib-other}. In the bibliography one finds several kinds of coupling which are more or less physically plausible. Among them, the interaction terms which are linear combinations of the dark sector components of the cosmic mixture, are well known \cite{luis-roy}. The following interaction has been investigated, for instance, in the references \cite{brasileiros, luis-roy},

\bea Q_1=3H(\alpha\,\rho_m+\sigma\,\rho_\vphi),\label{1-i-term}\eea where $\alpha$, $\sigma$ are adjustable constant parameters. The particular case when $\alpha=\sigma$ has been explored in \cite{jakubi}, while the case $\sigma=0$ was studied in \cite{coley} (see also \cite{saridakis}). The energy exchange imposed by the interaction term

\bea Q_2=3(\Gamma_m\,\rho_m+\Gamma_\vphi\,\rho_\vphi),\label{2-i-term}\eea where $\Gamma_m$, $\Gamma_\vphi$ are the constant transfer rates, is motivated by similar model of reheating \cite{turner}, curvaton decay \cite{malik}, and the decay of dark matter into radiation \cite{cen}. Many more complex (non linear) combinations have been also investigated (see, for instance, Ref. \cite{chen, saridakis}).

The question now is: are any imaginable kinds of coupling acceptable provided they are useful to solve a given cosmological problem, or do a given kind of non minimal coupling have to pass a compatibility check? If so, what can be an adequate check of compatibility? Since a non minimal coupling of matter can be correlated with an additional non gravitational interaction, it seems to us quite natural to require that any kind of coupling designed to solve a given problem; either within a cosmological setting, or in other contexts, has to be compatible with a fundamental theory of physics.

What comes out from requiring that the above interaction terms (\ref{1-i-term}) and (\ref{2-i-term}) be compatible with a fundamental theory of the interactions? In the next section we shall seek an answer to this question by assuming the string theory to be such a fundamental theory.

\section{The compatibility check}\label{consistency}

In the present paper, under the requirement of compatibility with the string effective theory -- what we call as the ``compatibility check'' -- we mean that a given interaction term $Q_X=Q_X(H,\rho_m,\dot\vphi,V)$ can be matched with the one in Eq. (\ref{string-i-term}), i. e. 

\bea Q_X(H,\rho_m,\dot\vphi,V)=\beta\,\dot\vphi\,\rho_m/2.\label{constraint}\eea 

In a cosmological setting the above expression means an additional constraint on the field variables $a(t)$ or $H:=\dot a/a$, $\rho_m(t)$, $\vphi(t)$, and $V(\vphi)$, which are involved in the cosmological equations.\footnote{Here we consider a particular FRW cosmological model fueled by a mixture of two fluids: (i) a pressureless dust fluid with energy density $\rho_m$ (dark matter), and (ii) an effective scalar field fluid with energy and parametric pressure given by $$\rho_\vphi=\dot\vphi^2/2+V(\vphi),\;p_\vphi=\dot\vphi^2/2-V(\vphi),$$ respectively.} Given the Einstein-scalar field equations with an additional interaction term included,

\bea &&3H^2=\rho_m+\frac{\dot\vphi^2}{2}+V(\vphi),\nonumber\\
&&\dot\rho_m+3H\rho_m=Q,\;\ddot\vphi+3H\dot\vphi=-\frac{dV}{d\vphi}-\frac{Q}{\dot\vphi},\label{cosmo-feqs}\eea there are three differential equations (\ref{cosmo-feqs}) to determine four unknown functions $a(t)$, $\rho_m(t)$, $\vphi(t)$ and $V(\vphi(t))$, i. e., the system of equations (\ref{cosmo-feqs}) is not closed. Recall that the interaction term $Q$ is an a priori given function $Q(H,\rho_m,\dot\vphi,V)$. From the above it follows that the equation (\ref{constraint}) is just an additional constraint on the field variables which closes the system of equations (\ref{cosmo-feqs}), and allows us, in principle, to solve it. 

In what follows we shall apply the above criterion to test the compatibility of the interacting models driven by (\ref{1-i-term}) and (\ref{2-i-term}) with the string effective theory. Since the general case is always difficult to handle, here we shall focus in several particular solutions which are of cosmological importance.

\subsection{de Sitter expansion}\label{de-sitter}

For a de Sitter solution [$\dot H=0\;\Rightarrow\;H=H_0$], independent on whether there is additional interaction or not, from the Friedmann equation in (\ref{cosmo-feqs}), and the derived equation $2\dot H=-\rho_m-\dot\vphi^2$, one gets

\bea 3H^2_0=\rho_m+\dot\vphi^2/2+V,\;\rho_m=-\dot\vphi^2.\nonumber\eea Given that only non-negative $\rho_m$ and $\dot\vphi^2$ are to be considered, the right-hand equality above is satisfied only if 

\bea \rho_m=\dot\vphi=0\;\Rightarrow\;&&3H_0^2=V(\vphi_0)=V_0,\nonumber\\
&&Q=\beta\,\rho_m\,\dot\vphi/2=0.\nonumber\eea In other words, in a stage of de Sitter expansion the non gravitational interactions turn off.

Hence, if require that the coupling depicted by the interaction term $Q_1$ in (\ref{1-i-term}) be compatible with the string effective theory, then $Q_1=3\sigma\,H_0 V_0=0$, which means that the de Sitter solution exists only for the particular kind of the non gravitational interaction $Q_1$ when $\sigma=0$ \cite{saridakis, coley}. A similar analysis, this time for the kind of interaction driven by $Q_2$ in Eq. (\ref{2-i-term}), leads to concluding that, if make this coupling to pass the compatibility check, then the de Sitter solution exists only if $$\Gamma_\vphi=0\;\Rightarrow\;Q_2=3\Gamma_m\,\rho_m.$$

In general, for models of non gravitational interactions which are compatible with a fundamental theory of physics, the de Sitter solution exists if the interaction term $Q_X$ can be factored by the energy density of the dark matter; 

\bea Q_X\propto\rho_m\,f(H,\rho_m,\dot\vphi,V).\label{ds-int}\eea Examples of this kind of interaction are $Q_X=3\alpha\,H\rho_m$, and also \cite{saridakis, chen}: 

\bea &&Q_X=\Gamma\,\rho_m,\;Q_X=\alpha_0 H^{3-2n}\,\rho_m^n,\nonumber\\
&&Q_X=\beta_0 \kappa^{2n} H^{1-2n}\,\rho_m^n\,\dot\vphi^2,\nonumber\eea where $\alpha_0$, $\beta_0$, $n$, $\kappa^2$ and $\Gamma$ are adjustable constants.

\subsection{Matter scaling solution}\label{scaling-sol}

If one wants to make the interaction term (\ref{1-i-term}) compatible with the one derived from the string effective theory in Eq. (\ref{string-i-term}), i. e., if require that $Q_1$ be written as $Q_1=\beta\dot\vphi\rho_m/2$, then one obtains the following relationship between $\dot\vphi$ and the Hubble parameter:

\bea \dot\vphi=6k(1+n\,r)H,\;r:=\frac{\rho_\vphi}{\rho_m},\;k:=\frac{\alpha}{\beta},\;n:=\frac{\sigma}{\alpha},\label{eq1}\eea where we assumed that $\alpha\neq 0$, and $\rho_m\neq 0$. The former relationship can be written in the form of the following quadrature:

\bea \vphi(t)=6k\int (1+n\,r) H dt+\vphi_0.\label{eq1'}\eea 

For a scaling solution [$r=r_0$] the above integral can be straightforwardly computed

\bea \vphi(t)=6k(1+n\,r_0)\ln a(t)+\vphi_0,\label{scaling-sol-1}\eea having the matter dominated solution as a particular case [$r=0$, $n\neq 0$] $\vphi(t)=6k\ln a(t)+\vphi_0$. This solution is also obtained if consider the particular coupling when $\sigma=0$ ($\Rightarrow$ $n=0$) \cite{coley}. Solutions of the kind (\ref{scaling-sol-1}) are found within the context of vacuum Einstein-scalar field system (no coupling) for the particular situation with vanishing potential $V(\vphi)=0$ \cite{dario}.

Since the general situation [$r=r(t)$] is very difficult to handle, in what follows, due to its role in cosmology, we shall focus in the scaling solution with $r=r_0$. In this case Eq. (\ref{eq1}) implies that

\bea &&\dot\vphi=6\bar k\,H\;\Rightarrow\;a(\vphi)=a_0\,e^{\vphi/6\bar k},\nonumber\\
&&\bar k:=k\,(1+n\,r_0),\;a_0:=e^{-\vphi_0/6\bar k},\label{eq1''}\eea i. e., $a(t)$ and $\vphi(t)$ are not independent of each other any more. If substitute $\dot\vphi$ ($\ddot\vphi$) from Eq. (\ref{eq1''}) into the Klein-Gordon equation (\ref{cosmo-feqs}) one obtains $$\dot H+3H^2=-\frac{1}{6\bar k}\frac{dV}{d\vphi}-\frac{Q/H}{(6\bar k)^2},$$ which, when compared with the Raychaudhury equation $$\dot H+3H^2=\rho_m/2+V,$$ yields the following first order differential equation for the self-interaction potential

\bea -\frac{dV}{d\vphi}=\left(3\bar k+\frac{\beta}{2}\right)\,\rho_m+6\bar k\,V,\label{edo-pot-1}\eea where we have taken into account $Q=\beta\,\dot\vphi\,\rho_m/2=3\beta\bar k\,H\,\rho_m$. The continuity equation for the matter degree of freedom $$\dot\rho_m=-3H\rho_m+Q=-3H(1-\beta\bar k)\rho_m,$$ implies that $$\rho_m(\vphi)=\rho_0\,a^{-3(1-\beta\bar k)}=M_0\,e^{-(1-\beta\bar k)\vphi/2\bar k},$$ where $M_0:=\rho_0\,a_0^{-3(1-\beta\bar k)}$. Eq. (\ref{edo-pot-1}) can then be written in the following way:

\bea -\frac{dV}{d\vphi}=\bar\beta\,e^{-\mu\vphi}+6\bar k\,V,\label{edo-pot-1'}\eea where $\bar\beta:=(3\bar k+\beta/2)\,M_0$, $\mu:=(1-\beta\bar k)/2\bar k$. This equation is easily integrated to obtain a self-interaction potential which is a combination of exponentials:

\bea V(\vphi)=V_{01}\,e^{-6\bar k\vphi}+V_{02}\,e^{-\mu\vphi},\label{pot-vphi-1}\eea where $V_{01}$ is an integration constant, and $V_{02}:=\bar\beta/(\mu-6\bar k)$. Hence, if require compatibility of the interaction term $Q_1$ in Eq. (\ref{1-i-term}) -- arbitrary $\alpha\neq 0$, $\sigma\neq 0$ -- with the string effective theory, the matter-scaling solution is driven by a potential of the form (\ref{pot-vphi-1}) exclusively. Meanwhile, for the particular case with $\sigma=0$ \cite{saridakis, coley}, the potential $V(\vphi)$ in Eq. (\ref{pot-vphi-1}) (replace $\bar k\rightarrow k$) is the only type of potential which makes the interaction term $Q_1$ to be compatible with a fundamental theory of physics, no matter which solution we were exploring.

Let us assume now it is the interaction term (\ref{2-i-term}) which is required to be matched with (\ref{string-i-term}): $Q_2=\beta\,\dot\vphi\,\rho_m/2$. As before the above equality means a constraint on the field variables which reduces the number of degrees of freedom. Actually, the mentioned constraint, in general, amounts to the following relationship:

\bea \dot\vphi_\pm=\frac{\beta\rho_m}{6\Gamma_\vphi}\pm\sqrt{\frac{\beta^2\rho_m^2}{36\Gamma_\vphi^2}-2\frac{\Gamma_m}{\Gamma_\vphi}\rho_m-2V},\label{eq2}\eea which means that $\dot\vphi_\pm=\dot\vphi_\pm(\rho_m,V)$, i. e., there is one less degree of freedom. 

The constraint $Q_2=\beta\,\dot\vphi\,\rho_m/2$ can be written, alternatively, in the following form:

\bea &&\dot\vphi=6K(1+N\,r),\;K:=\frac{\Gamma_m}{\beta},\;N:=\frac{\Gamma_\vphi}{\Gamma_m},\nonumber\\
&&\vphi(t)=6K\int (1+N\,r)\,dt+\vphi_0.\label{eq2'}\eea For a scaling solution [$r=r_0$] one gets

\bea \vphi(t)=6K(1+N\,r_0)\,t+\vphi_0.\label{scaling-sol-2}\eea Since, for this solution 

\bea \dot\vphi=6K\,(1+N\,r_0)=\sigma_0=const.,\label{eq2''}\eea then, one can straightforwardly integrate the continuity equation for the matter to get $\rho_m=M_0\,e^{\beta\vphi/2}/a^3$, where $M_0$ is an integration constant. Besides, equations (\ref{eq2}) and (\ref{eq2''}) lead to

\bea V=\left(\frac{\beta\sigma_0}{6\Gamma_\vphi}-\frac{\Gamma_m}{\Gamma_\vphi}\right)\,\rho_m-\frac{\sigma_0^2}{2},\nonumber\eea which, together with the Friedmann equation in (\ref{cosmo-feqs}), yield

\bea a^3(\vphi)=\frac{12M_0(1+\xi_0)}{\sigma_0^2\beta^2}\,e^{\beta\vphi/2}\left(1+\zeta_0\,e^{-\beta\vphi/4}\right)^2,\nonumber\eea where $B_0$ is an integration constant, and $$\xi_0:=\frac{\beta\sigma_0}{6\Gamma_\vphi}-\frac{\Gamma_m}{\Gamma_\vphi},\;\zeta_0:=\frac{\beta\sigma_0 B_0}{\sqrt{12M_0(1+\xi_0)}}.$$ One ends up with the following potential

\bea V(\vphi)=\frac{\beta^2\sigma_0^2\xi_0}{12(1+\xi_0)}\,\frac{e^{\beta\vphi/2}}{(e^{\beta\vphi/4}+\zeta_0)^2}-\frac{\sigma_0^2}{2}.\label{pot-vphi-2}\eea 

It follows a similar conclusion than in the former case: If require the interaction term $Q_2$ in Eq. (\ref{2-i-term}) to pass the compatibility check, the matter-scaling solution is driven by a potential of the form (\ref{pot-vphi-2}) exclusively.

\subsection{Matter domination solution}\label{mat-dom-sol}

A solution characterized by matter domination, in the context of the cosmological model we are exploring in this paper, means that $$\rho_\vphi=0\;\Rightarrow\;\dot\vphi=0,\;V(\vphi)=0.$$ This entails the constraint $$Q_X(H,\rho_m,\dot\vphi,V)=\beta\,\dot\vphi\,\rho_m/2=0.$$ Hence, if one looks for a matter-dominated solution within a cosmological model with a coupling of the form $Q_1=3H(\alpha\rho_m+\sigma\rho_\vphi)$, since at matter domination $\rho_\vphi=0$, and since compatibility with string theory means $Q_1=0\;\Rightarrow\;\alpha=0$, then, only an interaction term of the form $Q_1=3\sigma\,H\rho_\vphi$ would be compatible with a fundamental theory of physics. The same applies to $Q_2$ in Eq. (\ref{2-i-term}), this time $Q_2=3\Gamma_\vphi\rho_\vphi$ would be the only admissible possibility.

\subsection{Summary of partial results}\label{summary}

Due to their importance for further discussion (see Sec. \ref{discussion}), here we summarize the most relevant results of this section. What we did in this section was to match a couple of well known interaction terms (\ref{1-i-term}), (\ref{2-i-term}), with the only kind of coupling which arises naturally in the Einstein's frame formulation of the string effective theory  [$Q=\beta\,\dot\vphi\,\rho_m/2$] Eq. (\ref{string-i-term}), and to look for the consequences of the additional constraint implied by the matching. The match is required by compatibility with a fundamental theory of physics (the string theory in the present case). We called this as the ``compatibility check'' for the Einstein-scalar system. The main results forced by this test are:

\begin{enumerate}

\item Existence of the de Sitter solution is possible only for couplings $Q_X$ of the form $$Q_X\propto\rho_m\,f(H,\rho_m,\dot\vphi,V).$$ This means that for the interacting functions $Q_1$ (\ref{1-i-term}), and $Q_2$ (\ref{2-i-term}), the de Sitter solution does not exist in the general case [$\alpha,\sigma\neq 0$, $\Gamma_m,\Gamma_\vphi\neq 0$], but only if $Q_1=3\alpha\,H\rho_m$, and $Q_2=3\Gamma_m\rho_m$. Recall that, if relax the compatibility requirement, the coupled Einstein-scalar system does actually admit de Sitter solutions \cite{luis-roy}.

\item Under the requirement of compatibility of the coupled Einstein-scalar system with the string effective theory, the matter-scaling solution exists if the self-interaction potential $$V(\vphi)=V_{01}\,e^{-6\bar k\vphi}+V_{02}\,e^{-\mu\vphi},$$ for the case where $Q_1=3H(\alpha\rho_m+\sigma\rho_\vphi)$, and $$V(\vphi)=\frac{\beta^2\sigma_0^2\xi_0}{12(1+\xi_0)}\,\frac{e^{\beta\vphi/2}}{(e^{\beta\vphi/4}+\zeta_0)^2}-\frac{\sigma_0^2}{2},$$ if $Q_2=3(\Gamma_m\rho_m+\Gamma_\vphi\rho_\vphi)$.

\item If the coupled Einstein-scalar system (\ref{cosmo-feqs}) is required to pass the compatibility check, then the matter domination solution does not exist in the general case neither for $Q_1=3H(\alpha\rho_m+\sigma\rho_\vphi)$, nor for $Q_2=3(\Gamma_m\rho_m+\Gamma_\vphi\rho_\vphi)$ [$\alpha,\sigma\neq 0$, $\Gamma_m,\Gamma_\vphi\neq 0$]. Only for the particular kinds of interaction $Q_1=3\sigma\,H\rho_\vphi$, $Q_2=3\Gamma_\vphi\rho_\vphi$, the matter-dominated solution exists under the mentioned requirement.

\end{enumerate}

\section{Chameleon Coupling}\label{chameleon-coupling}

As mentioned in the introduction, a way to avoid the tight constraints on additional non gravitational interactions \cite{5-force, will, baessler} can be through the chameleon effect \cite{khoury, cham-brax-prd, cham-tsujikawa-jcap, cham-takami-prd, cham-salehi-prd, cham-polarski-prd, cham-banerjee-prd}. Regrettably, in order to get such a desirable effect several ad hoc assumptions are required which may be unjustified:

\begin{enumerate} 

\item Consideration of self-interaction potentials of the runaway type only, which should be monotonically decreasing functions and should fulfill a number of assumptions on the asymptotic behavior of the ratio of their derivatives \cite{khoury, cham-polarski-prd}. 

\item Geometry approximated by Minkowski space \cite{khoury, cham-takami-prd, cham-polarski-prd}.

\item A certain ``energy density which, in a cosmological setting would be conserved'' in the Einstein's frame, is commonly considered \cite{khoury}, instead of the energy density which is measured by co-moving observers, or the one which is conserved in the Jordan's (string) frame. 

\end{enumerate} 

In this section we shall approach this subject backwards: First we relax the assumption 2 above and assume a relativistic background metric \cite{cham-tsujikawa-jcap}, then we write the resulting Einstein's equations which, under the assumption of pressureless background matter, can be written as a first order differential equation for the self-interaction potential. Finally, we make reasonable assumptions on the kinds of admissible matter energy density as a function of the chameleon field $\vphi$. The functional form of the potential is then uniquely singlet out as a result of the above procedure.

Let us start with the total action $S_{\text{tot}}$, with $S_\vphi^E$ and $S_m^E$ given by Eq. (\ref{ef-actions}) or, equivalently, with the derived field equations (\ref{ef-feqs}) and the continuity Eq. (\ref{m-cont-eq}). I. e., we shall be considering a single matter species. Additionally, it will be assumed that the matter is in the form of a perfect fluid with stress-energy tensor 

\bea T^{(m)}_{\mu\nu}=(\rho+p)\,u_\mu u_\nu+p\,g_{\mu\nu},\label{matt-set}\eea where $\rho$ is the energy density of the fluid as measured by a co-moving observer, $p$ its pressure, and $$u^\mu=\left(-\frac{1}{\sqrt{-g_{00}}},0,0,0\right),$$ is the 4-velocity field of the fluid. We shall assume, also, a static, spherically symmetric spacetime which, in spherical coordinates $t,r,\theta,\psi$ is depicted by the line-element:

\bea ds^2=-e^{\xi(r)}dt^2+e^{\zeta(r)}dr^2+r^2d\Omega^2,\label{line-e}\eea where $d\Omega^2=d\theta^2+\sin^2\theta\,d\psi^2$. 

Under these assumptions the field equations (\ref{ef-feqs}) and the continuity equation (\ref{m-cont-eq}) read

\bea &&\xi''+\frac{\xi'^2}{2}-\frac{\xi'\zeta'}{2}+\frac{2\xi'}{r}=e^\zeta\left(\rho+3p-2V\right),\nonumber\\
&&-\xi''-\frac{\xi'^2}{2}+\frac{\xi'\zeta'}{2}+\frac{2\zeta'}{r}=e^\zeta\left(\rho-p+2V\right)+2\vphi'^2,\nonumber\\
&&\frac{2(e^\zeta-1)}{r^2}+\frac{\zeta'-\xi'}{r}=e^\zeta\left(\rho-p+2V\right),\nonumber\\
&&\vphi''+\left(\frac{\xi'-\zeta'}{2}+\frac{2}{r}\right)\vphi'=e^\zeta\left[\frac{\beta}{2}(\rho-3p)+\frac{dV}{d\vphi}\right],\nonumber\\
&&p'+\frac{\xi'}{2}\left(\rho+p\right)=\frac{\beta}{2}(3p-\rho)\,\vphi',\label{spher-feqs}\eea where the tilde denotes derivative with respect to the $r$-coordinate, and, in place of the standard form of Einstein's equations $G_{\mu\nu}=T_{\mu\nu}$, we have used the alternative form $R_{\mu\nu}=T_{\mu\nu}-g_{\mu\nu}\,T/2$. Of these only 4 are independent equations on the six unknown field variables $\xi,\zeta,\vphi,V,\rho,p$. This means that we are free to impose two additional conditions on the unknown field variables (see below). Note also that the interaction term (\ref{i-tensor}) is given by $$Q_\mu=\left(0,Q_r,0,0\right),\;Q_r=-\frac{\beta}{2}(3p-\rho)\,\vphi'.$$

By combining the first two equations above one obtains,

\bea \frac{\xi'+\zeta'}{r}=e^\zeta(\rho+p)+\vphi'^2.\label{combi-1}\eea 

Let us further assume that the spherically symmetric fluid behaves like dust, i. e., it is a non-relativistic fluid $p/\rho\ll 1$, so that, to a good approximation we can set $p'=p=0$. In this case, from the last equation in (\ref{spher-feqs}) -- see Eq. (15) of Ref. \cite{cham-tsujikawa-jcap} -- it follows that: 

\bea \xi'=-\beta\vphi'.\label{rel-1}\eea 

As it will be immediately shown in Eq. (\ref{pot-eq}), given a known functional dependence of energy density $\rho=\rho(\vphi)$, the equation (\ref{rel-1}) represents a constraint on the form of the self-interaction potential $V(\vphi)$ which is compatible with the Einstein's equations (\ref{spher-feqs}). This means that, if specify how $\rho$ depends on $\vphi$ as it is usually done \cite{khoury}, we are not free neither to choose the potential, nor to impose ad hoc constraints on it. This relationship is still valid in the weak field and small velocities limit, and, even if in the weak field, small velocities and low energy density limit it becomes an identity, it is still a valid constraint on the kinds of potential which are compatible with the given field equations. In spite of its simplicity and potential implications, the role of the latter equation is usually underestimated. 

By substituting (\ref{rel-1}) back into the first equation in (\ref{spher-feqs}), and then comparing with the 4th (Klein-Gordon) equation, one obtains the following first-order ordinary differential equation for the self-interaction potential:

\bea \beta\frac{dV}{d\vphi}=2V-\left(1+\frac{\beta^2}{2}\right)\,\rho.\label{pot-eq}\eea 

Depending on the way $\rho$ depends on $\vphi$ this equation may be exactly solved to get $V=V(\vphi)$ explicitly. There are, at least, three reasonable possibilities $\rho=\rho(\vphi)$ which deserve consideration.

\subsection{String frame $\rho_\text{SF}$ does not depend on $\vphi$}

Perhaps the most natural choice is to assume that in the string frame the matter energy density $\rho_\text{SF}$ does not depend explicitly on the dilaton $\vphi$ as long as $\rho_\text{SF}$ and $\rho_\phi$ are separately conserved in this frame. In this case, in the EF, due to the coupling of the dilaton the matter energy density (measured by a co-moving observer) does actually depend on $\vphi$:

\bea \rho=\rho(\vphi,r)=e^{2\beta\vphi}\,\rho_\text{SF}(r).\label{case-1}\eea If substitute $\rho(\vphi)$ in this equation back into (\ref{pot-eq}), after straightforward integration, one obtains

\bea V(\vphi)=V_{01}\,e^{2\beta\vphi}+V_{02}\,e^{2\vphi/\beta},\label{pot-1}\eea where the constant $$V_{01}=\left(\frac{2+\beta^2}{1-\beta^2}\right)\frac{\rho_\text{SF}}{4},$$ and $V_{02}$ is an arbitrary integration constant.

\subsection{Einstein's frame conserved $\rho^{\text{cons}}_\text{EF}$}

This is the most frequently studied case within the framework of chameleon models \cite{khoury, cham-brax-prd, cham-tsujikawa-jcap, cham-takami-prd, cham-salehi-prd, cham-polarski-prd}. Here the quantity which, within a cosmological setting is conserved in the Einstein's frame [$\dot\rho^{\text{cons}}_\text{EF}+3H\rho^{\text{cons}}_\text{EF}=0$] is $\rho^{\text{cons}}_\text{EF}=\rho\,\exp(-\beta\vphi/2)$, i. e., in the present case\footnote{In the original bibliography \cite{khoury}, due to different choice of units and of Lagrangian density, $\rho=e^{3\beta\vphi/M_\text{pl}}\,\rho^{\text{cons}}_\text{EF}(r)$ is obtained instead of Eq. (\ref{case-2}).}

\bea \rho=\rho(\vphi,r)=e^{\beta\vphi/2}\,\rho^{\text{cons}}_\text{EF}(r).\label{case-2}\eea After substituting $\rho$ from (\ref{case-2}) back into Eq.(\ref{pot-eq}), it is not complicated to obtain the following solution $V(\vphi)$:

\bea V(\vphi)=\bar V_{01}\,e^{\beta\vphi/2}+\bar V_{02}\,e^{2\vphi/\beta},\label{pot-2}\eea where $$\bar V_{01}=\left(\frac{2+\beta^2}{4-\beta^2}\right)\,\rho^{\text{cons}}_\text{EF},$$ and $\bar V_{02}$ is an arbitrary integration constant.

\subsection{Co-moving energy density does not depend on $\vphi$}

In this case it is assumed that the energy density $\rho$ in (\ref{pot-eq}) does not depend explicitly on $\vphi$; $\rho=\rho_\text{C}(r)$. After this assumption one gets the following potential:

\bea V(\vphi)=V_0+C_0\,e^{2\vphi/\beta},\;V_0:=\frac{1}{2}\left(1+\frac{\beta^2}{2}\right)\,\rho_\text{C},\label{pot-3}\eea where $C_0$ is an arbitrary integration constant.

\subsection{Summary of this section}

Neither of the potentials (\ref{pot-1}), (\ref{pot-2}), and (\ref{pot-3}), are of the runaway type, since none of them fulfills the necessary asymptotics \cite{khoury}:

\bea &&\lim_{\vphi\rightarrow\infty}V=0,\;\lim_{\vphi\rightarrow\infty}\frac{V_{,\vphi}}{V}=0,\;\lim_{\vphi\rightarrow\infty}\frac{V_{,\vphi\vphi}}{V_{,\vphi}}=0,\;...,\nonumber\\
&&\lim_{\vphi\rightarrow 0}V=\infty,\;\lim_{\vphi\rightarrow 0}\frac{V_{,\vphi}}{V}=\infty,\;\lim_{\vphi\rightarrow 0}\frac{V_{,\vphi\vphi}}{V_{,\vphi}}=\infty,\;...\nonumber\eea Actually, the potentials (\ref{pot-1}) and (\ref{pot-2}) are of the general form

\bea V(\vphi)=A\,e^{\beta_1\vphi}+B\,e^{\beta_2\vphi},\label{pot-gen}\eea where $A$, $B$ are constants (these can be, in general, explicit functions of $r$), and $\beta_1=2\beta$, $\beta_2=2/\beta$ for the former, while $\beta_1=\beta/2$, $\beta_2=2/\beta$ for the latter. Hence, depending on whether $\beta$ is a large quantity or a small one, either

\bea \lim_{\vphi\rightarrow\infty}V\propto A\,e^{\beta_1\vphi},\;\lim_{\vphi\rightarrow\infty}\frac{V_{,\vphi}}{V}=\lim_{\vphi\rightarrow\infty}\frac{V_{,\vphi\vphi}}{V_{,\vphi}}=\beta_1,\nonumber\eea or

\bea \lim_{\vphi\rightarrow\infty}V\propto B\,e^{\beta_2\vphi},\;\lim_{\vphi\rightarrow\infty}\frac{V_{,\vphi}}{V}=\lim_{\vphi\rightarrow\infty}\frac{V_{,\vphi\vphi}}{V_{,\vphi}}=\beta_2.\nonumber\eea In both cases

\bea &&\lim_{\vphi\rightarrow 0}V=A+B,\;\lim_{\vphi\rightarrow 0}\frac{V_{,\vphi}}{V}=\frac{\beta_1 A+\beta_2 B}{A+B},\nonumber\\
&&\lim_{\vphi\rightarrow 0}\frac{V_{,\vphi\vphi}}{V_{,\vphi}}=\frac{\beta^2_1 A+\beta^2_2 B}{\beta_1 A+\beta_2 B}.\nonumber\eea For the third potential (\ref{pot-3}) one has 

\bea \lim_{\vphi\rightarrow\infty}V\propto C_0\,e^{2\vphi/\beta}\sim\infty,\;\lim_{\vphi\rightarrow 0}V=V_0,\nonumber\eea

\bea &&\lim_{\vphi\rightarrow\infty}\frac{V_{,\vphi}}{V}=\lim_{\vphi\rightarrow 0}\frac{V_{,\vphi}}{V}=\frac{2C_0}{\beta},\nonumber\\
&&\lim_{\vphi\rightarrow\infty}\frac{V_{,\vphi\vphi}}{V_{,\vphi}}=\lim_{\vphi\rightarrow 0}\frac{V_{,\vphi\vphi}}{V_{,\vphi}}=\frac{4C_0^2}{\beta^2}.\nonumber\eea

The main conclusion drawn from this study is that, whenever the following assumptions are satisfied: (i) the field equations (\ref{spher-feqs}) -- in particular the last equation -- govern the dynamics of matter,\footnote{This includes the weak field limit of equations (\ref{spher-feqs}).} and (ii) the approximation $p'=p=0$ is valid, i. e., whenever Eq. (\ref{rel-1}) is satisfied, we are not free to impose ad hoc constraints on the self-interaction potential $V$ since, once $\rho=\rho(\vphi)$ is specified, the derived equation (\ref{pot-eq}) uniquely determines the functional form of $V=V(\vphi)$. As a matter of fact, if consider reasonable functions $\rho=\rho(\vphi)$ -- in particular the assumption $\rho\propto e^{\beta\vphi/2}$ usually undertaken in the bibliography \cite{khoury, cham-brax-prd, cham-tsujikawa-jcap, cham-takami-prd, cham-salehi-prd, cham-polarski-prd} -- the allowed self-interaction potentials are of the general form (\ref{pot-gen}). Hence, potentials of the runaway form \cite{khoury}, such as, for instance, the power-law potential $$V(\vphi)=M^{4+n}\vphi^{-n},$$ where $n$ is a positive constant and $M$ has units of mass, are not allowed.

\section{Discussion}\label{discussion}

Here we shall first to discuss on the results of our study regarding the chameleon behavior of scalar fields, and then we shall thoroughly discuss the results on the compatibility of given matter couplings of cosmological importance, with the string effective theory. Although the former results come from local (static, spherically symmetric) exploration of the Einstein-scalar system, while the latter ones are derived from their cosmological inspection, both studies are tightly correlated: Testing of the compatibility of several well known cosmological interaction terms with a fundamental theory of physics places theoretical constraints on the kind of such couplings which are allowed, meanwhile, reviewing of the chameleon effect after relaxing several common, perhaps unjustified ad hoc assumptions, poses reasonable doubts on the most elegant way in which the non minimal coupling of matter can avoid the tight constraints coming from five-force/equivalence principle experiments \cite{5-force, will, baessler}.

\subsection{Chameleon behavior}

In the weak gravity, small velocities limit [$\xi,\,\zeta,\,\xi',\,\zeta'\propto{\cal O}(\xi^2)\sim{\cal O}(\zeta^2)$], in the approximation $p'=p=0$, the Einsteins's equations (\ref{spher-feqs}) become

\bea &&\xi''+\frac{2\xi'}{r}=(1+\zeta)(\rho-2V),\;\frac{\zeta'+\xi'}{r}=(1+\zeta)\,\rho,\nonumber\\
&&\frac{\zeta'-\xi'}{r}+\frac{2\zeta}{r^2}=(1+\zeta)(\rho+2V),\nonumber\\
&&\vphi''+\frac{2}{r}\,\vphi'=(1+\zeta)\left(\frac{\beta}{2}\,\rho+\frac{dV}{d\vphi}\right),\;\xi'=-\beta\vphi',\label{wfl-feqs}\eea where we have ignored quantities $\sim\xi^2\sim\zeta^2\sim\xi'^2\sim\zeta'^2\sim\vphi'^2\sim\xi'\zeta'\sim\xi'\vphi'...$, and smaller. Hence, Eq. (\ref{pot-eq}) $$\beta\frac{dV}{d\vphi}=2V-\left(1+\frac{\beta^2}{2}\right)\,\rho,$$ is still valid, as it is a direct consequence of the equations (\ref{wfl-feqs}). This means that, even in the weak field/low velocities limit of the Einstein-scalar system (\ref{spher-feqs}), the results of Sec. \ref{chameleon-coupling} are still valid.

An important ingredient of the chameleon behavior is that the chameleon dynamics is not governed by $V(\vphi)$, but by an effective self-interaction potential which is the sum of $V(\vphi)$ and a density dependent term: $$V_\text{eff}(\vphi)=V(\vphi)+\rho(\vphi)=V(\vphi)+e^{\beta_1\vphi}\rho_0,$$ where, for the third case studied above $\beta_1=0$, and $\rho_0$ is an energy density which can be an explicit function of $r$, but not of $\vphi$. Besides, it is required that $V_\text{eff}$ has a minimum \cite{khoury}. The value of the mass of small fluctuations about the minimum $\mu^2_{\vphi,\text{min}}$, depends on the energy density of the environment [$\rho_0=\rho_0(r)$] in such a way that, the larger $\rho_0$ the larger the mass of the chameleon. For the cases of interest considered here one obtains (compare with equations (\ref{pot-1}), (\ref{pot-2}), and (\ref{pot-3}))

\bea &&V_\text{eff}(\vphi)=\tilde{V}_{01}\,e^{2\beta\vphi}+V_{02}\,e^{2\vphi/\beta},\;\tilde{V}_{01}=V_{01}+\rho_\text{SF},\nonumber\\
&&V_\text{eff}(\vphi)=\tilde{\bar V}_{01}\,e^{\beta\vphi/2}+\bar V_{02}\,e^{2\vphi/\beta},\;\tilde{\bar V}_{01}=\bar V_{01}+\rho^\text{cons}_\text{EF},\nonumber\\
&&V_\text{eff}(\vphi)=\tilde V_0+C_0\,e^{2\vphi/\beta},\;\tilde V_0=V_0+\rho_\text{C},\nonumber\eea respectively, i. e., $V_\text{eff}$ has no minimum. Then, regrettably, whenever the assumptions undertaken in the section \ref{chameleon-coupling} of this paper are fulfilled, there is no room for chameleon behavior.

Interestingly, if forget about the ad hoc assumptions made on the self-interaction potential, the study of Ref. \cite{khoury} is perfectly compatible with our assumptions here: (i) weak field/small velocities limit of (\ref{spher-feqs}), and (ii) non relativistic (pressureless) fluid with energy density $\rho$. The discussion above then poses reasonable doubt on the validity of the chameleon effect as we know it.

\subsection{Matter couplings which are compatible with a fundamental theory of physics}

Now we turn to the cosmological aspect of non minimal coupling of matter. In section \ref{consistency} we have approached the compatibility of several kinds of matter coupling with the string effective theory, by matching given interaction terms (\ref{1-i-term}), (\ref{2-i-term}), with the one which is straightforwardly derived from string theory (\ref{string-i-term}), within a FRW cosmological setting. The matching (\ref{constraint}) $$Q_X(H,\rho_m,\dot\vphi,V)=\beta\,\dot\vphi\,\rho_m/2,$$ amounts to an additional constraint on the field variables $H$, $\rho_m$, $\vphi$, and $V$, which closes the system of equations (\ref{cosmo-feqs}). Take as an example the particular kind of interaction \cite{coley, saridakis}

\bea Q_X(H,\rho_m,\dot\vphi,V):=3\alpha\,H\rho_m.\label{coley-case}\eea Then, the matching (\ref{constraint}) means that 

\bea \beta\dot\vphi=6\alpha\,H\;\Rightarrow\;a(\vphi)=a_0\,e^{\beta\vphi/6\alpha},\label{a-vphi}\eea where $A$ is an arbitrary integration constant. If substitute back (\ref{a-vphi}) into (\ref{cosmo-feqs}) one obtains

\bea &&-\beta\frac{dV}{d\vphi}=\nu\rho_m+6\alpha V,\;\rho_m=M_0\,e^{-\mu\vphi},\nonumber\\
&&\nu:=3\alpha+\frac{\beta^2}{2},\;\mu:=\frac{\beta(1-\alpha)}{2\alpha},\;M_0:=\frac{\rho_0}{a_0^{3(1-\alpha)}},\nonumber\eea or, after integrating

\bea V(\vphi)=V_{01}\,e^{-\frac{6\alpha}{\beta}\,\vphi}+V_{02}\,e^{-\mu\vphi},\label{pot-disc}\eea where $V_{01}$ is an integration constant, and $$V_{02}:=\frac{(6\alpha+\beta^2)M_0}{2\beta\mu-12\alpha}.$$ If consider the inverse relationship $(a/a_0)^\lambda=e^{\vphi}$ [$\lambda:=6\alpha/\beta$], the Friedmann equation in (\ref{cosmo-feqs}) can be put in the form of the following quadrature:

\bea \int\frac{dz}{\sqrt{A+B\,z^{2(1-\lambda/\mu)}}}=\pm\frac{\mu\lambda\,(t-t_0)}{2\sqrt{3(1-\lambda/6)}},\label{quad}\eea where $z:=a^{\mu\lambda/2}$, $A:=(M_0+V_{02})a_0^{\mu\lambda}$, $B:=V_{01}a_0^{\lambda^2}$, and $t_0$ is another integration constant. Once the values of the free parameters $\alpha$, $\beta$ ($\mu$, $\lambda$) are specified, the integral above can be straightforwardly computed to obtain $t=t(z)$, or after reversal, $z=z(t)$ $\Rightarrow$ $a=a(t)$.

Another way around is to specify the cosmological dynamics (properly, to specify a particular solution of the cosmological equations), then the functional form of the self-interaction potential $V=V(\vphi)$ is also (uniquely) specified. Since, the most general situation (no specification of the dynamics) is very difficult to handle, in Sec. \ref{consistency} we focused in three exact solutions which are of prime importance in cosmological settings: (i) de Sitter expansion, (ii) matter scaling solution, and (iii) matter domination solution. The first solution above is important since the present stage of the cosmic expansion is quite well approximated by the $\Lambda$-cold dark matter ($\Lambda$CDM) model, which is characterized by a de Sitter attractor \cite{de-models}. A matter scaling attractor, when correlated with accelerated expansion, is useful to overcome the cosmic coincidence problem \cite{wands, amendola, interaction, brasileiros, luis-roy}, while matter domination is a necessary (transient) stage of the cosmic evolution, indispensable to explain the amount of cosmic structure we see \cite{amendola, avelino}.

The results of Sec. \ref{consistency} indicate that (see Sec. \ref{summary} for a brief summary of results), in the general case [$\alpha,\sigma,\Gamma_m,\Gamma_\vphi$], if require compatibility with the string effective theory, none of the interaction terms (\ref{1-i-term}), (\ref{2-i-term}) allow for the existence of the de Sitter and the matter-dominated solutions at once. This is in obvious contradiction with the accepted cosmological paradigm since the current stage of the cosmic evolution is very close to a de Sitter expansion, while the matter-dominated solution is of prime importance for the right amount of cosmic structure to be formed. For particular cases of (\ref{1-i-term}) and (\ref{2-i-term}) the above mentioned solutions can exist but exclusively: the de Sitter solution exists if either $Q_X=3\alpha H\rho_m$ or $Q_X=3\Gamma_m\rho_m$, while the matter domination solution exists if either $Q_X=3\sigma H\rho_\vphi$ or $Q_X=3\Gamma_\vphi\rho_\vphi$. However, once the compatibility requirement is relaxed the situation gets normal and the Einstein-scalar system of cosmological equations (\ref{cosmo-feqs}) with $Q$ given by either (\ref{1-i-term}) or (\ref{2-i-term}), admits the mentioned solutions. 

From the mathematical point of view it might be unnatural to require compatibility of couplings (\ref{1-i-term}) and (\ref{2-i-term}) with the string theory, since this entails an additional constraint on the field equations which cuts off an important subset from their space of solutions. On the other hand, since non minimal coupling of matter can be associated with an additional non gravitational interaction, it seems natural to require that such a coupling is to be derived from (at least compatible with) a fundamental theory of the interactions. The results just discussed suggest that matter couplings other than the exponential (dilaton) one -- particularly those given by (\ref{1-i-term}) and (\ref{2-i-term}) -- should be dismissed. This is not a conclusive result since several stringent assumptions have been made. Among them, the starting ones: (i) matter couplings are to be derived from a fundamental theory of physics, and (ii) string theory is such a fundamental theory.

\subsection{Lagrangian matter couplings}

Under ``Lagrangian matter coupling'' we understand a matter coupling which is derivable from or compatible with a Lagrangian principle. Since the coupling (\ref{lag-i-term}) represents the most general Lagrangian formulation of an additional non-gravitational interaction of matter with a scalar field, we shall say that a given coupling $Q_X=Q_X(H,\rho_m,\dot\vphi,V)$ is compatible with a Lagrangian formulation if the following constraint is fulfilled:

\bea Q_X(H,\rho_m,\dot\vphi,V)=\frac{1}{2}\frac{d\ln\chi}{d\vphi}\,\dot\vphi\,\rho_m.\label{lag-consistency}\eea 

As with the issue of compatibility with the string effective theory, from the mathematical point of view, the additional constraint (\ref{lag-consistency}) on the field variables entails that a set of (families of) solutions is cut off from the space of solutions of the field equations. If take a look at the analysis in section \ref{consistency}, one sees that the results 1 and 3 in \ref{summary} are independent of whether $Q=\beta\dot\vphi\rho_m/2$ or $Q=(d\ln\sqrt\chi/d\vphi)\dot\vphi\rho_m$. Following the line of reasoning above, one is then forced to conclude that the couplings (\ref{1-i-term}) and (\ref{2-i-term}) are not compatible with a Lagrangian picture. Given the power of Lagrangians this can be, perhaps, a more robust argument against the cosmological interaction terms of the kind explored here.

\section{Conclusion}\label{conclusion}

In the present paper we approached the matter coupling issue with the hope to seek for additional feasible criteria that might help in establishing the validity of several types of interaction terms which are useful in cosmological setups. We have assumed that one such criterion can be to check the compatibility of given types of coupling with a fundamental theory of physics: the string effective theory. The results of our study, although not conclusive, suggest that matter couplings other than the exponential dilaton coupling are very unlikely to arise in cosmological scenarios. Same conclusion is obtained if one seeks for compatibility with a Lagrangian picture. Our conclusions heavily rely on several stringent assumptions, among which the starting ones are: (i) matter couplings are to be derived either from a fundamental theory of physics or from a Lagrangian principle, and (ii) string theory is such a fundamental theory. Besides, for simplicity, our approach was based on the analysis of particular solutions which are useful in cosmology. 

We think more robust results require relaxation of some of the assumptions made here. Additionally, a detailed analysis of the phase space of the models studied in this paper, will lead to more conclusive results without the need of analytical solutions.
 
In this paper we have also reviewed the chameleon effect (an elegant result useful to overcome the stringent experimental constraints coming from five force and equivalence principle tests) by relaxing several of the assumptions which are usually considered in the bibliography on this subject. We have shown that under the following reasonable assumptions: (i) the weak field and low velocities limit is valid, and (ii) the background matter is in the form of a pressureless perfect fluid, the functional form of the chameleon self-interaction potential $V=V(\vphi)$ is fixed by the Einstein-scalar equations (\ref{spher-feqs})/(\ref{wfl-feqs}). This means that ad hoc assumptions on the asymptotics of the potential are unjustified. In particular, potentials of the runaway type are not allowed. This result raises reasonable doubt on the validity of the chameleon picture as we know it.

The author thanks SNI of Mexico and the "Programa PRO-SNI, Universidad de Guadalajara" for support under grant No 146912.

\end{document}